\documentclass[bibyear]{aa}
\usepackage{graphicx}
\usepackage{graphicx}
\usepackage{txfonts}
\usepackage{url}
\usepackage{hyperref}

\usepackage{natbib}
\usepackage{lineno}  
\usepackage{color}

\newcommand{\uvot}{\textit{Swift}/UVOT}
\newcommand{\nus}{\textit{Nu}STAR}
\newcommand{\xrt}{\textit{Swift}-XRT}

\newcommand{\po}{power-law}

\newcommand{\lp}{log-parabola}

\newcommand{\nh}{N$_H$}
\newcommand{\obj}{1ES\,0229+200}

\begin{document}

\title{Disentangling two spectral components in the X-ray emission of the  blazar \obj}
\titlerunning{Two spectral components of the  blazar \obj} 
\author{Alicja Wierzcholska\inst{1,2}
        \and
        Hubert Siejkowski\inst{3}
}
\authorrunning{A. Wierzcholska \& H. Siejkowski}

\institute{Institute of Nuclear Physics, Polish Academy of Sciences, ul. Radzikowskiego 152, 31-342 Krak\'{o}w, Poland \\
    \email{alicja.wierzcholska@ifj.edu.pl}
    \and Landessternwarte, Universit\"at Heidelberg, K\"onigstuhl 12, D 69117 Heidelberg, Germany
    \and Academic Computer Centre CYFRONET of the AGH University of Krakow, Nawojki 11, 30-950 Krak\'{o}w, P.O. Box 6, Poland
}

  \abstract{
X-ray observations are essential to achieve a deeper understanding of the broadband emission mechanism in blazars. Here, we present a long-term spectral and temporal analysis of X-ray and optical observations of 1ES\,0229+200 collected with the\textit{ Neil Gehrels Swift }Observatory from 2008 to 2024, complemented by hard X-ray observations from the Nuclear Spectroscopic Telescope Array (\textit{NuSTAR}).

The blazar 1ES\,0229+200 is a high-frequency, peaked BL Lac object, known for its exceptionally hard very high-energy (VHE) $\gamma$-ray spectrum extending up to 10\,TeV. 
In August 2021, \textit{NuSTAR} observed the source in a low X-ray state, revealing a concave spectral shape with a distinct upturn around 25\,keV. This feature contrasts with previous observations performed with \textit{NuSTAR} and \textit{Swift}-BAT, which showed no such spectral upturn.

Previous observations of 1ES\,0229+200 and broadband SED (spectral energy distribution)
modelling suggest that its X-ray emission extends beyond 100\,keV without a significant cutoff. The newly detected spectral upturn may indicate a transition between the synchrotron and inverse Compton components or could be linked to photohadronic processes involving high-energy neutrinos.

We discuss the implications of this finding in the context of blazar spectral energy distributions, particularly the potential existence of a third SED bump in the kiloelectronvolt to megaelectronvolt range. The observed spectral features support the hypothesis that 1ES\,0229+200 could be a source of high-energy neutrino emission.

} 
 

  \maketitle
\def\obj{1ES\,0229+200}

\section{Introduction}
Blazars are radio-loud active galactic nuclei characterised by relativistically beamed, non-thermal jet emission pointing towards the observer's line of sight \citep{Begelman, Urry}. 
These objects exhibit strong variability in both the spectral and temporal domains, with changes observable across all energy bands on timescales as short as days or even hours \citep[e.g.][]{WagnerWitzel}.
The broadband electromagnetic spectra of blazars feature two distinct peaks in the $\nu$ - $\nu F_\nu$ representation, one spanning from radio to optical-X-ray frequencies, and the other peaking at high-energy $\gamma$ rays.
The low-energy peak in the SED (spectral energy distribution) is usually explained by synchrotron radiation of relativistic electrons from the jet, whereas the high-energy component can be explained in leptonic and$/$or hadronic explanations \citep[see e.g.][]{Marsche, Bottcher}.
 
Lepto-hadronic models propose that hadrons are accelerated alongside leptons, resulting in processes such as pion production and synchrotron radiation from secondary particles \citep{Mannheim1993}.
A natural outcome of these secondary interactions is the production of neutrinos.
In such scenarios, neutrinos can be generated within optically thick regions, where the suppression of high-energy $\gamma$-ray emission ($E \sim \mathrm{GeV}$--$\mathrm{TeV}$) due to $\gamma\gamma$ absorption may lead to an enhanced flux in the soft-to-hard X-ray band \citep{Petropoulou2015, Murase2016, Reimer2019, Petropoulou2020, Oikonomou2021}.
X-ray observations thus provide key diagnostics of the emission processes at play and can contribute to the distinguishing between leptonic and hadronic origin models \citep[e.g.][]{Zhang2019}.

Blazars are classified in two subgroups: flat-spectrum radio quasars (FSRQs) and BL Lacertae (BL Lac) type objects.
BL Lac objects can further be categorized based on the location of the low-energy peak in their SED into high-energy-peaked (HBL), intermediate-energy-peaked (IBL), and low-energy-peaked (LBL) blazars \citep[e.g.][]{padovani95, Abdo2010}.
For LBL-type blazars the synchrotron bump is located in the infrared regime ($\nu_s \lesssim 10^{14}~\mathrm{Hz}$), for IBL-type blazars in the optical-UV range ($10^{14}~\mathrm{Hz} < \nu_s \lesssim 10^{15}~\mathrm{Hz}$), while for HBL blazars up to the X-ray domain \citep[$\nu_s > 10^{15}~\mathrm{Hz}$;][]{Abdo2010}.

As a result of these characteristics, the position of the X-ray spectrum within the broadband SED varies across the BL Lac subclasses. In HBL blazars, the X-ray spectrum generally covers the synchrotron domain, whereas in LBL blazars, it is located within the inverse Compton region of the SED. Interestingly, in some IBL blazars, the X-ray spectrum spans both the synchrotron and inverse Compton components, leading to a noticeable spectral upturn in the X-ray energy range visible for this type of blazar (see \citealt{Wierzcholska_swift} and references therein).

The high-frequency peaked BL Lac object \obj\ located at a redshift of $z = 0.1396$ \citep{Woo} is frequently observed at different energy wavelengths \citep[see e.g.][]{Rector, Giommi, Kaufmann, Wierzcholska_atom,  Cologna, Ehlert}.
The SED of \obj\ is reported to ranging even above 100\,keV without any significant cut-off \citep{Kaufmann}, classifying the blazar as an extreme HBL-type source. 
In the X-ray frequencies, it was discovered by the Einstein IPC Slew Survey \citep{Elvis}, and since then it has been the target of numerous X-ray observational campaigns.

\textit{NuSTAR} has clearly detected \object{1ES\,0229+200} up to energies of $\sim$50\,keV \citep{Furniss2015, Biteau2020, Sahakyan2021}. The X-ray spectrum in the \textit{NuSTAR} band is characterized by a very hard photon index, typically in the range $\Gamma \sim 1.5$--$2.0$, consistent with the continuation of the synchrotron component to very high energies. This hard X-ray tail is detected even during low flux states, suggesting a persistent population of ultra-relativistic electrons \citep{Sahakyan2021}.

The broadband spectral-energy-distribution modelling supports the interpretation that the X-ray emission in the \textit{NuSTAR} band arises from synchrotron radiation of highly energetic electrons. In particular, the extension of the synchrotron spectrum to hard X-rays, without any clear indication of a cut-off within the \textit{NuSTAR} band, places strong constraints on the maximum energy of accelerated electrons \citep{Furniss2015}. These electrons can also be responsible for the inverse Compton emission observed in the TeV regime, making \textit{NuSTAR} observations an essential component in one-zone synchrotron-self-Compton (SSC) modelling \citep{Aliu2014}.

Moreover, the hard TeV spectrum observed by H.E.S.S. and VERITAS \citep{Aharonian07, Aliu2014}, in combination with the hard X-ray data from \textit{NuSTAR}, has been used to place lower limits on the strength of the intergalactic magnetic field \citep[IGMF;][]{Tavecchio2010, Biteau2020}. These studies rely on the assumption of minimal variability and a hard intrinsic TeV spectrum, both of which are supported by the X-ray observations.

Despite being a bright optical and X-ray source, it is a rather faint radio and $\gamma$-ray emitter.
In the TeV $\gamma$ rays, it was detected with the H.E.S.S. telescopes in 2006 \citep{Aharonian07}.
Contrary to the X-ray variability, in the very high-energy gamma rays \obj\ exhibit hints of variability \citep[see e.g.][]{Aliu14}.

In this paper, we focus on the long-term X-ray variability of \obj\ using data collected with \xrt. We also report on the low-X-ray-state observations collected with \nus. 
The paper is organised as follows: Sect.~\ref{datanalysis} describes data used and  analysis techniques; results are given in Sect.~\ref{resluts}; while the studies are summarised in Sect.~\ref{summary}.

\section{Data analysis} \label{datanalysis}
\subsection{NuSTAR observations}
\nus\ is  a  Small  Explorer  satellite dedicated to X-ray observations in the energy regime of 3--79\,keV \citep{Harrison13}. Only
\obj\ was observed with \nus\ three times in 2013 and three times in 2021. 
The set of 2013 was already discussed by e.g. \cite{Pandey_nustar, Bhatta_nus, Wierzcholska_host}. 
Here, we focus on three observations taken in August 2021, corresponding to the ObsIDs of 10702609002, 10702609004, and 10702609006. 

For all three observations, the raw data were processed with the \nus\ Data Analysis Software package released within HEASoft (v.\,6.33.2). 
For the analysis, a source region was selected within a circle centred on the blazar. 
The same-size circular region, but located in a different position on a map,  was used in order to define background area.
 Different positions of the background region were used in order to check whether the background selection affects the analysis results. 
 In the analysis performed, only channels corresponding to the energy range of 3–40\,keV were selected. 
All spectra, along with the ancillary and response files, were generated using the \verb|nuproducts| task for point-like sources.
All spectra were binned using \verb|grppha| to have at least 20\,counts per channel.
For each ObsID, spectra were fitted using the power-law log-parabola (curved power-law). 
The models used are defined as follows: 

\begin{itemize}
 \item a single power law:
 \begin{equation}
\frac{dN}{dE}=N_p  \left( \frac{E}{E_0}\right)^{-{\gamma}},
\end{equation}
 with the spectral index $\gamma$ and the normalization $N_p$;

\item a log-parabola:
 \begin{equation}
\frac{dN}{dE}=N_l  \left( \frac{E}{E_0}\right)^{-({\alpha+\beta \log (E/E_0)})},
\end{equation}
 with the normalisation $N_l$, the spectral index $\alpha,$ and the curvature parameter $\beta$;

\item a broken power law:
\begin{equation}
\frac{dN}{dE} = N_b \times\left\{\begin{array}{ll} (E/E_b)^{-\Gamma_1} & \mbox{if $E < E_b$,}\\ (E/E_b)^{-\Gamma_2} & \mbox{otherwise,} \end{array}\right. 
\end{equation}
with the normalisation $N_b$, the spectral indices $\Gamma_1$ and $\Gamma_2,$ and the break energy $E_b$.
\end{itemize}

\noindent In the case of \po\ and \lp\ models, the scale energy $E_0$ is fixed at 1\,keV. 

A summary of all \nus\ observations taken both in 2013 and 2021 as well IDs of the corresponding IDs of simultaneous \xrt\ observations are given in Table~\ref{table_nustardata}. 
Table\,\ref{table_nustarallfits} includes results of the spectral fitting of the \nus\ data with three spectral models. 

\begin{table*}  
\caption[]{\textit{NuSTAR} observations of \obj\ and simultaneous \xrt ones.  }
\centering
\begin{tabular}{c|c|c|c|c}
\hline
\hline
 ObsID & Label  & Date  & Exposure (s) & XRT data ObsID   \\
           (1) &  (2) & (3) & (4) & (5) \\
\hline 
10702609006& nu21$\_$06  & 2021-08-13 05:11:09  & 56068  & 00031249108    \\
10702609004& nu21$\_$04  & 2021-08-11 04:51:09  & 58095  & 00031249106    \\
10702609002& nu21$\_$02  & 2021-08-08 05:51:09  & 95156  & 00031249101-00031249102     \\
60002047006& nu13$\_$06  & 2013-10-10 23:11:07  & 18021  & 00080245005    \\
60002047004& nu13$\_$04  & 2013-10-05 23:31:07  & 20289  & 00080245004     \\
60002047002& nu13$\_$02  & 2013-10-02 00:06:07  & 16256  & 00080245003    \\               
\hline
\hline
\end{tabular}
\tablefoot{The columns present (1) observation ID of \textit{NuSTAR} observations; (2) label of \nus\ observations, used in Fig.~\ref{fig:nustar_spectra}; (3) date of NuSTAR observations; (4) Exposure of \textit{NuSTAR} observations, given in seconds; (5) ID of the corresponding, simultaneous \xrt\ observations. }
\label{table_nustardata}
\end{table*}

\begin{table*}  
\caption[]{Spectral fits parameters for the \nus\ observations of \obj.}
\centering
\begin{tabular}{c|c|c|c|c|c|c}
\hline
\hline
\nus\ observation &  Model & Normalization & $\gamma$ or $\alpha$ or $\Gamma_1$ & $\beta$ or $\Gamma_2$ &  $E_{br}$ & $\chi^2$(d.o.f.)  \\
 (1) &  (2) & (3) & (4) &  (5) &  (6) & (7)  \\
\hline 
nu21$\_$02  & PL        & 4.0$\pm$0.2 & 2.32$\pm$0.02 & -- & --  &  138(149)  \\   
            & LP        & 2.2$\pm$0.4 & 1.7$\pm$0.2 &0.4$\pm$0.1  & --  & 135(148)   \\  
            & BKNPL     & 4.0$\pm$0.2 & 2.32$\pm$0.02 & 1.51$\pm$0.02 & 26$\pm$2  &  135(148)  \\ 
\hline
nu21$\_$04  & PL        & 4.4$\pm$0.3 & 2.40$\pm$0.03  & -- & --  &   115(93)   \\   
            & LP        & 1.6$\pm$0.4 & 1.3$\pm$0.3 &0.6$\pm$0.1   & --  & 100(92)   \\  
            & BKNPL     & 4.4$\pm$0.3 & 2.41$\pm$0.03  & 1.25$\pm$0.02 & 26$\pm$2  &  98(92)  \\ 
\hline
nu21$\_$06  & PL        & 5.3$\pm$0.3 & 2.37$\pm$0.03  & -- & --  & 100(101)    \\    
            & LP        & 2.9$\pm$0.6 & 1.7$\pm$0.2 & 1.7$\pm$0.2  & --  &  96(100)  \\  
            & BKNPL     & 5.3$\pm$0.3 & 2.37$\pm$0.03 &1.61$\pm$0.03  & 25$\pm$2  &  98(100)  \\  
\hline 
nu13$\_$02  & PL        & 7.9$\pm$0.5 &2.02$\pm$0.03 & -- & --  & 117(142)  \\
            & LP        &  5.1$\pm$1.1  & 1.6$\pm$0.2  & 0.2$\pm$0.1 & --  & 111(141)   \\
 
\hline
nu13$\_$04  & PL       & 8.8$\pm$0.4    & 2.03$\pm$0.02   & -- & --  &   228(193)   \\
            & LP       & 5.7$\pm$0.9   & 1.6$\pm$0.2  & 0.2$\pm$0.1    & --  & 219( 192)   \\
 
\hline
nu13$\_$06  & PL        &  7.7$\pm$0.4   & 1.95$\pm$0.02   & -- & --  & 174( 183)    \\
            & LP        &  6.0$\pm$1.0  &  1.7$\pm$0.2  & 0.1$\pm$0.1  & --  &  171(182)  \\
 
\hline
\hline
\end{tabular}
\tablefoot{The following columns present (1) observation ID;   (2) fitted model PL (single power-law), LP (logparabola), or BKNPL (broken \po); (3) normalisation in 10$^{-3}$\,erg\,cm$^{-2}$\,s$^{-1}$;  (4) spectral parameter $\gamma$, $\alpha,$ or $\Gamma_1$; (5) curvature parameter in the case of \lp\ model or second spectral index in the case of broken \po\ model; (6) break energy of the broken \po\ model; (8) $\chi^2$ and number of degrees of freedom for a model used.}
\label{table_nustarallfits}
\end{table*}

\subsection{Swift-XRT and Swift-UVOT observations}
The observations collected in the period of 2008-2024, corresponding to the ObsIDs of 00031249001 - 00016432016, were analysed with the HEASoft package (v.\,6.33.2). 
All events were cleaned and calibrated using the \verb|xrtpipeline| task, and all PC- and WT-mode data in the energy range of 0.3-10\,keV were used. 
For the spectral fitting, the data were grouped with the \verb|grappha| task in order to have a minimum of 20 counts per bin.
The spectra were fitted with the single-power-law model together with the Galactic-column-density value of 1.16 $\cdot$ 10$^{21}$ cm$^{-2}$ \citep{Willingale}  using the \verb|XSPEC| software \citep{Arnaud96}. 
For all fits, the \nh\ was frozen as a fixed parameter. 
The results of the spectral fitting are listed in Table\,\ref{table_xrtfluxes}.

Simultaneously, the blazar was observed with the UVOT instrument on board \textit{Swift} in  the U\,(345\,nm), B\,(439\,nm), and V\,(544\,nm) filters. 
For all observations corresponding to the ObsIDs of 00031249001-00016432016, the instrumental magnitudes were calculated using \verb|uvotsource| including all photons from a circular region with a radius of 5''.
The background was determined from a circular region with a radius of 10'' near the source region that is not contaminated with signal from nearby sources. 
The flux-conversion factors are provided by \mbox{\cite{Poole08}}.
All data were corrected for or the dust absorption using the reddening $E(B-V) = 0.1185$~mag as provided by \cite{Schlafly} and the ratios of extinction to reddening, $A_{\lambda} / E(B-V)$, for each filter provided by \cite{Giommi06} are used. 

We note here that for the last epoch of the XRT observations, the corresponding UVOT points are not available. 
During this epoch, a technical issue of the instrument and an increase in noise in one of the three onboard gyroscopes were reported. 
The UV and optical observations for this period cannot be analysed \citep[for details, see][]{Cenko}.

In addition, Table\,\ref{table_xrtfluxes} includes the spectral fit parameters for a power-law fit of the \xrt\ observations simultaneous to six \nus\ observations. 
In the case of the \nus\ observation with ObsID of 10702609002, two \xrt\ observations are merged.

\begin{table*}  
\caption[]{Spectral fits parameters for \xrt\ observations of \obj. }
\centering
\begin{tabular}{c|c|c|c}
\hline
\hline
 ObsIDs & $N$  & $\gamma$ & $\chi^2$(d.o.f.) \\
           (1) &  (2) & (3) & (4) \\
\hline 
00031249108 &       2.45$\pm$0.11   &  2.03$\pm$0.06     &   110(115)   \\
00031249106 &       1.65$\pm$0.19   &  1.56$\pm$0.15  &            26(28)       \\
00031249101-00031249102 &      2.22$\pm$0.14   &  1.99$\pm$0.09  &       56(65)  \\
00080245005  &       4.31$\pm$0.35   &  1.69$\pm$0.10  &                   42(46)  \\
00080245004  &        4.34$\pm$0.14   & 1.73$\pm$0.04  &                190(227)    \\
00080245003 &         3.59$\pm$0.22   & 1.86$\pm$0.08  &                 92(81)    \\

\hline
\hline
\end{tabular}e
\tablefoot{The columns present (1) ID of observations;   (2)  normalisation in 10$^{-3}$\,erg\,cm$^{-2}$\,s$^{-1}$;  (3) spectral parameter $\gamma$; (4)  $\chi^2$ and number of degrees of freedom for a model used.}
\label{table_xrtfluxes}
\end{table*}

\section{Results} \label{resluts}

\subsection{Characterisation of the variability}

\begin{figure*}
\centering{\includegraphics[width=0.9\textwidth]{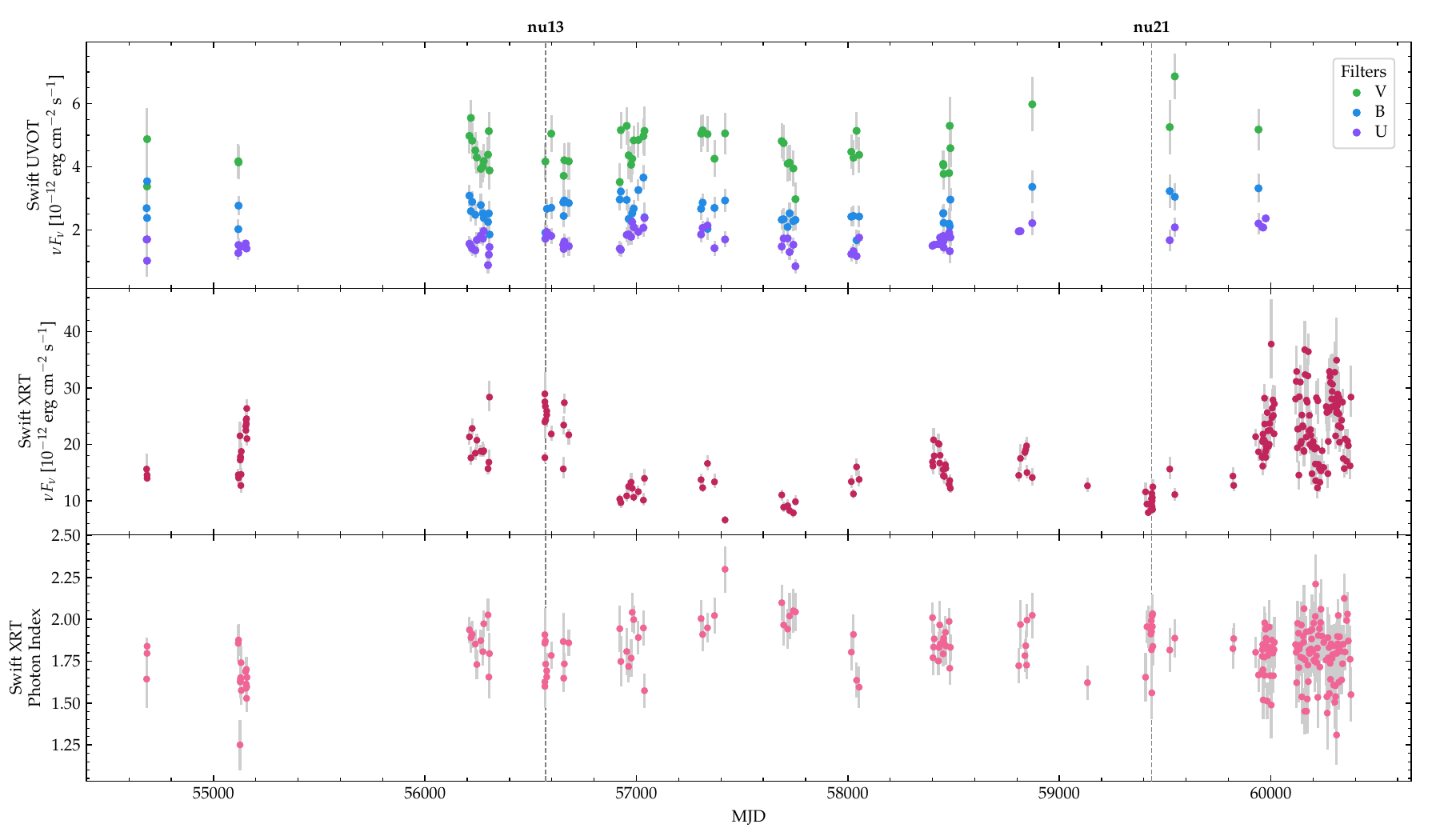}}
\caption{Multi-wavelength light curve of \obj\ presenting long-term (2008-2024) observations of the blazar.  The panels present optical \textit{Swift}-UVOT observation in U, B, V filters; X-ray (0.3-10\,keV) flux; and corresponding \po\ photon index. Two epochs of \nus\ observations taken in 2013 and 2021 are marked with vertical lines and noted as nu13 and nu21, respectively. }
\label{fig:mwl}
\end{figure*}

The long-term light curve of \obj\ spanning from 2008 to 2024, which includes both optical and X-ray observations, is shown in Fig. \ref{fig:mwl}.
The figure is divided into three panels that display the variability in optical fluxes, X-ray flux, and the photon index, the latter two derived from \po\ spectral fits within the 0.3-10\,keV energy band. The X-ray data reveal significant flux variability, with changes of a factor of four, accompanied by shifts in the photon index ranging between 1.3 and 2.2.

The temporal variability observed in the light curve can be quantified using the fractional variability amplitude, as defined by \cite{Vaughan03} and \cite{Poutanen08} as

\begin{equation}
 F_{var}= \sqrt{\frac{S^2-e^2}{F^2}},
\end{equation}
where $S^2$ is the variance, $e^2$ is the mean square error, and $F$ is the mean flux. 
The uncertainties of $F_{var}$ are calculated following the formula by \cite{Poutanen08}:

\begin{equation}
  \delta  F_{var}= \sqrt{F_{var}^2+(\sigma^2)} -F_{var},
\end{equation}
with the error in the normalised excess variance $\sigma$ given as \citep{Vaughan03}
\begin{equation}
 \sigma = \sqrt{\left( \sqrt{\frac{2}{N} }\frac{e^2}{F^2} \right)^2   + \left( \sqrt{\frac{e^2}{N}}\frac{2 F_{var}}{F} \right)^2        },
\end{equation}
where $N$ is the number of data points in the light curve.

Consistent with previous findings for \obj, the long-term variability in the X-ray band (0.3-10\,keV) is more pronounced than in the optical bands. The X-ray light curve exhibits considerable variability, with long-term flux variations of a factor of three, while the optical V, B, and U bands show only marginal variability. This corresponds to fractional variability amplitudes ($F_{var}$) of $31 \% \pm 1 \%$ for the X-ray band and around $(5-12) \% \pm 2 \% $ for the optical frequencies.

Due to the limited number of \nus\ observations, it is not feasible to determine the fractional variability amplitude for these data. However, for each \nus\ observation, the fluxes in the 3-10\,keV and 3-40\,keV energy bands were calculated and are summarised in Table \ref{table_nustarfluxes}. These flux measurements confirm a significantly lower X-ray flux level in the 2021 observations compared to the 2013 data.
The lower X-ray-flux level is also notable for the \xrt\ observations of the blazar taken for the same period. 
Furthermore, in the case of the 2021 \nus\ observation of \obj\, notable flux variability between individual observations within both energy bands is evident, confirming daily variability also in the hard X rays.

\begin{table}  
\caption[]{Summary of X-ray fluxes.}
\centering
\begin{tabular}{c|c|c}
\hline
\hline
 ObsID & Flux 3-10\,keV  & Flux 3-40\,keV \\
           (1) &  (2) & (3) \\
\hline 
nu21$\_$06   &  5.46$\pm$0.33 & 9.37$\pm$0.85  \\
nu21$\_$04   &  4.30$\pm$0.27 & 7.26$\pm$0.55  \\
nu21$\_$02   &  4.51$\pm$0.27 & 7.94$\pm$0.56  \\
nu13$\_$06   &  14.3$\pm$1.2 & 28.0$\pm$1.8  \\
nu13$\_$04   &  15.8$\pm$1.2 & 30.8$\pm$2.0  \\
nu13$\_$02   &  16.0$\pm$1.1 & 32.2$\pm$1.9  \\               
\hline
\hline
\end{tabular}
\tablefoot{X-ray fluxes measured in the energy bands of 3-10\,keV (2) and 3-40\,keV (3) for all six NuSTAR observations. All values given in 10$^{-12}$\,erg\,cm$^{-2}$\,s$^{-1}$.}
\label{table_nustarfluxes}
\end{table}

\subsection{Correlations}

\begin{figure*}
\centering{\includegraphics[width=0.35\textwidth]{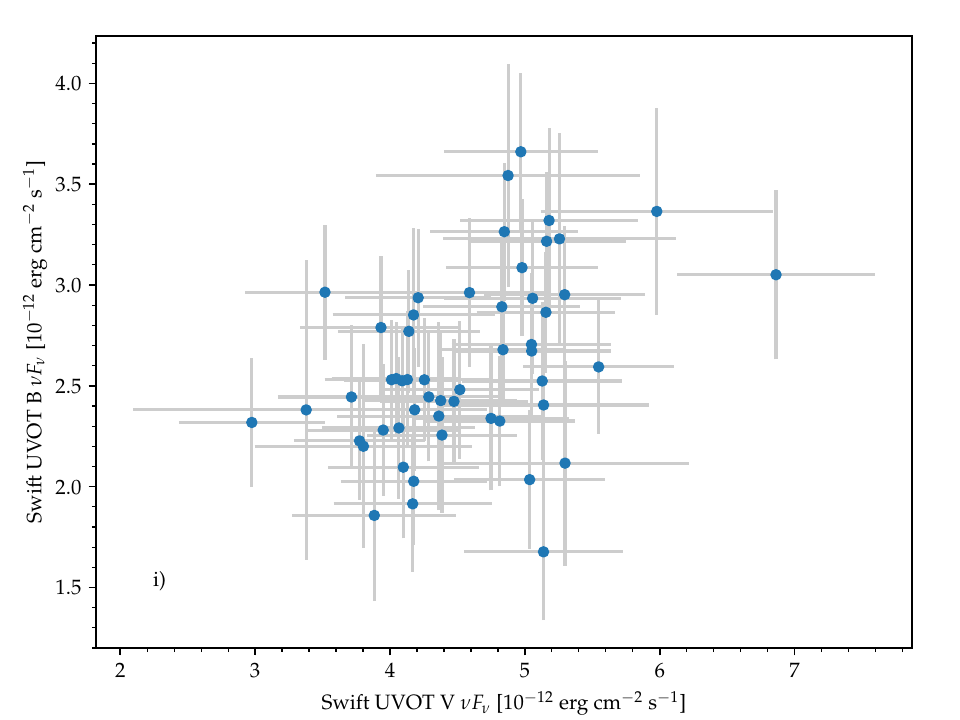}}
\centering{\includegraphics[width=0.35\textwidth]{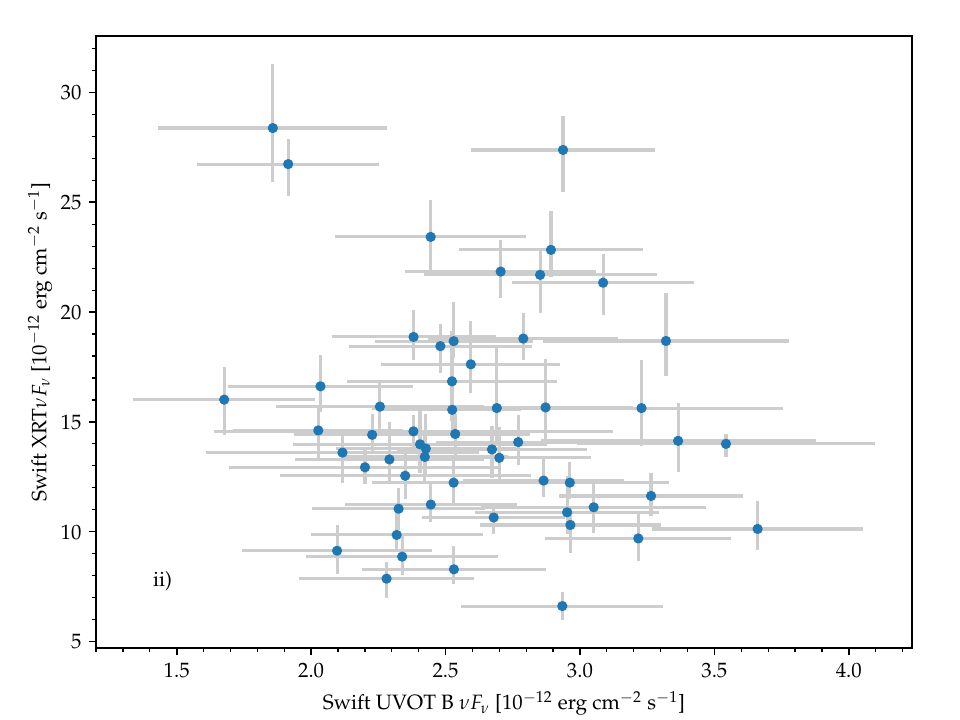}} \\
\centering{\includegraphics[width=0.35\textwidth]{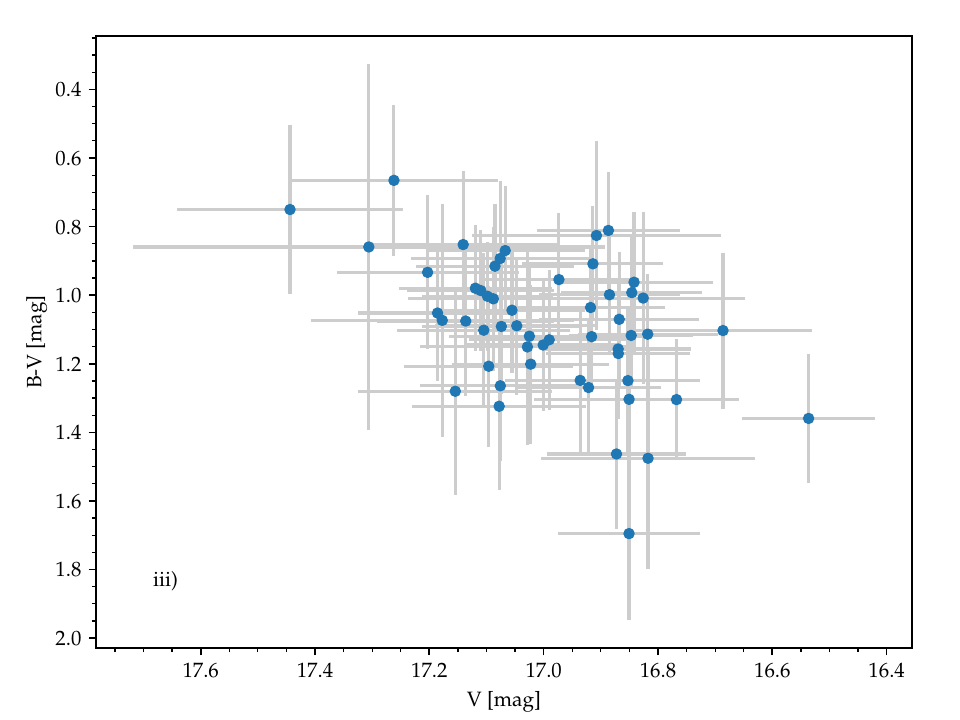}}
\centering{\includegraphics[width=0.35\textwidth]{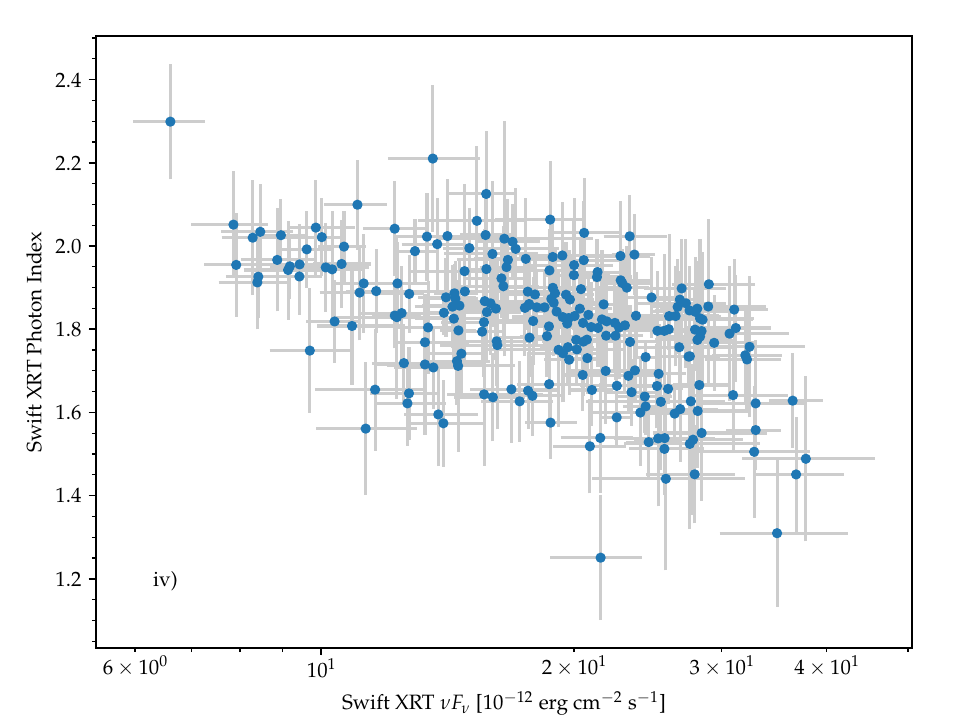}}
\caption{Correlation plots for \obj. The following figures present flux-flux comparison of \uvot\ V and B filter; flux-flux comparison of \uvot\ V flux and \xrt\ flux measured in the energy range of 0.3-10\,keV; colour-magnitude diagram for V magnitude and $B-V$ colour; flux-index comparison for \xrt\ observations.}
\label{fig:corr}
\end{figure*}

Simultaneous X-ray and optical observations of \obj\ were performed with \xrt\ and \nus,\, respectively. 
Fig.\,\ref{fig:corr} shows comparison of of the optical and X-ray fluxes.
That includes a flux-flux comparison for the \uvot\ B and V filters; the optical B filter and \xrt\ 0.3--10\,keV flux; the colour-magnitude diagram for optical V and B filters; and a comparison of the X-ray 0.3--10\,keV flux and corresponding photon index.
For all cases,  each point in a plot corresponds to a single \textit{Swift} observation, and only simultaneous observations (with the same ObsID) are considered here.

The main results of the correlation analysis are listed below.
\begin{itemize}
    \item[i)] The optical \uvot\ B and V fluxes show a slight correlation, with a Pearson correlation coefficient of 0.5$\pm$0.1\footnote{Errors for the Pearson correlation coefficients are calculated as described by \cite{Wierzcholska_0048}.}. The relation is weak, which can be attributed to several factors, including the lack of correction for host-galaxy contamination and large uncertainties due to the short exposures used in the comparison.
    
    \item[ii)] Simultaneous X-ray and optical observations do not reveal any apparent correlation. The Pearson correlation coefficient for this relation is $-$0.13$\pm$0.09, indicating an absence of any significant trend. The lack of correlation between optical and X-ray fluxes probably results from the low variability in the optical band compared to the significant flux changes observed in X-rays (see e.g. Fig.~\ref{fig:mwl}).
    
    \item[iii)] The colour--magnitude diagram shows a clear correlation, with a coefficient of 0.5$\pm$0.1. Although still a weak trend, it can be explained by the same effects noted in (i). This bluer-when-brighter behaviour has previously been reported for \obj\ based on long-term optical monitoring \citep[e.g.][]{Wierzcholska_atom}.
    
    \item[iv)] A negative correlation is observed between the X-ray flux and photon index, with a Pearson coefficient of 0.52$\pm$0.04, indicating a harder-when-brighter trend in the 0.3--10\,keV X-ray band.
\end{itemize}

\subsection{Spectral properties}

\begin{figure*}
\centering{\includegraphics[width=0.98\textwidth]{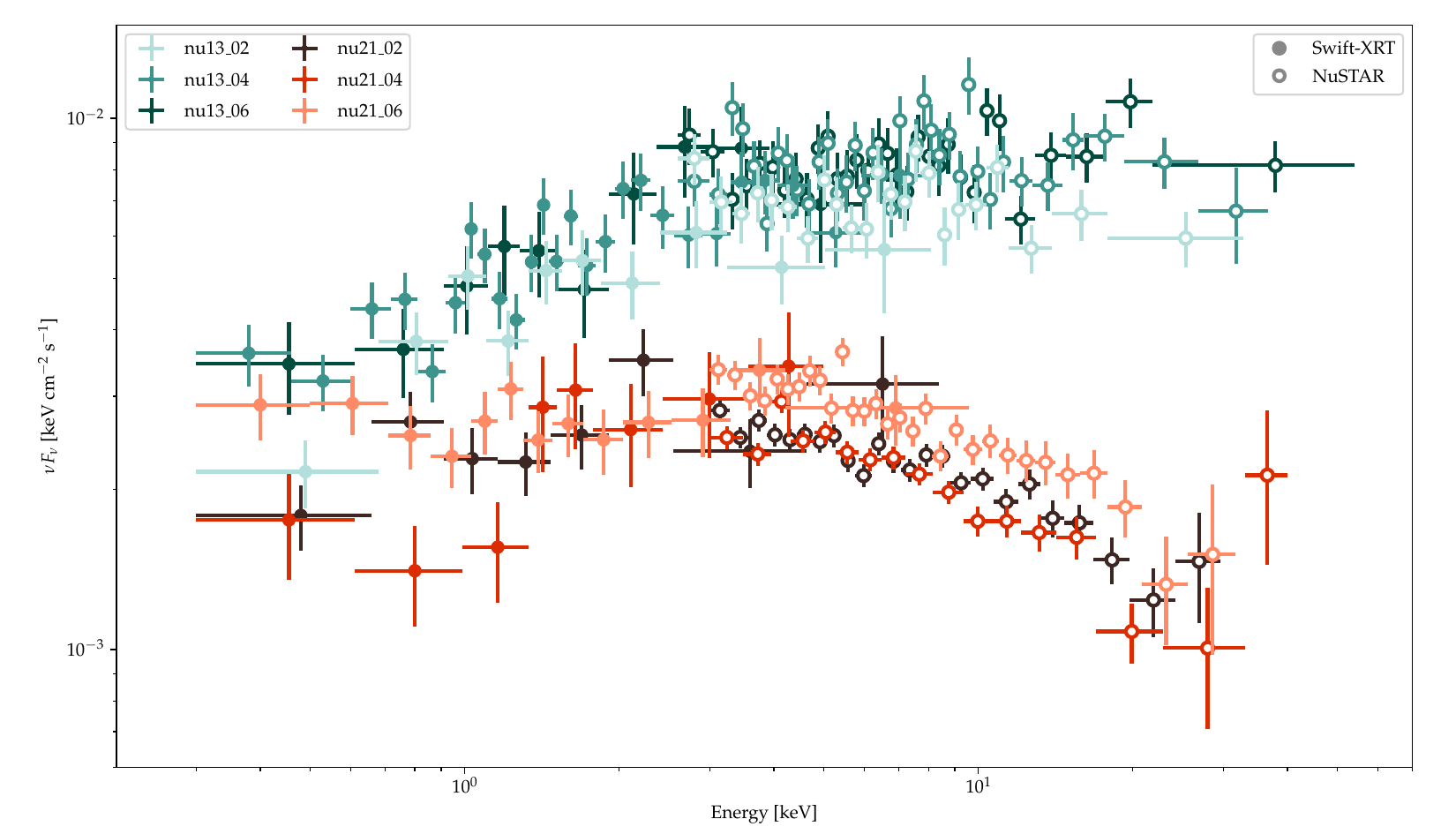}}
\caption{Spectral energy distributions presenting \textit{NuSTAR} and \xrt\ observations for six epochs (3 in 2013 and 2021). \xrt\ observations are corrected for the influence of the hydrogen Galactic column density with \nh\ provided by \cite{Willingale}.  The last two digits of the label denote the ObsIDs of the \nus\ observations. The \xrt\  and \textit{NuSTAR} observations are distinguished by filled and hollow markers, respectively. The full ObsIDs are listed in Table \ref{table_nustardata}. In the case of 2013 \nus\ data, spectral points are adopted from \cite{Wierzcholska_host}. }. 
\label{fig:nustar_spectra}
\end{figure*}

In the spectral analysis, we focused on the 2021 \nus\ data (ObsID 10702609002-10702609006) and the corresponding simultaneous \xrt\ observations.
The details of the simultaneous \xrt\ data used are listed in Table \ref{table_nustardata}. All \nus\ spectra were fitted using three different models: a single \po, \lp, and a broken \po, each including the Galactic column density as provided by \cite{Willingale}. Table\,\ref{table_nustarallfits} presents the resulting fitting parameters for all tested models for the \nus\ observations, while Table\,\ref{table_xrtfluxes} presents all spectral parameters for the \po\ fits to the X-ray data. 
The errors for all parameters are quoted at the 1-$\sigma$ confidence level.
For comparison, Table \ref{table_nustarallfits} also includes the spectral parameters of the \po\ and \lp\ fits to the 2013 \nus\ data.
Since there is no clear preference for a curved spectral shape in this case, the parameters of the broken \po\ model are not included in the table.

Figure \ref{fig:nustar_spectra} shows the \nus\ spectra for different epochs of \nus\ observations of the blazar, from both 2013 and 2021. To provide a wider range of X-ray spectral data for \obj, simultaneous \xrt\ observations are also included for all \nus\ observations.
In the case of the X-ray spectral presented in Fig.\,\ref{fig:nustar_spectra}, the \xrt\ and \nus\ data points are the results of the separate fitting of the observations taken with these two instruments. 
For all three observations, the $\chi^2$ values and spectral shapes of the 2021 \nus\ spectra, visible in Fig. \ref{fig:nustar_spectra}, indicate a hard X-ray excess above 25\,keV.
Such a spectral feature was not detected in the \nus\ observations of the blazar performed in 2013.

To compare the goodness of fit among the three tested models, we employed the F-test \cite[e.g.][]{statystyka}. 
This allowed us to assess whether adding an extra parameter to a more complex model---accounting for either curvature or a spectral break---provides a statistically significant improvement over the simpler single-power-law model. 
The test was applied to all three \nus\ observations obtained in 2021. 
The hard X-ray excess is most clearly visible in the observation labelled nu21$\_$04, and this is confirmed by the F-test results, which indicate that the inclusion of the additional parameter in the more complex model yields a highly significant improvement in the fit, at a confidence level exceeding $4\sigma$ ($p \ll 0.001$). 
For observations  nu21$\_$02  and  nu21$\_$06, the improvement is still present but less pronounced, reaching only the 1-2$\sigma$ level ($p \approx 0.1$).

\section{Discussion} \label{summary}
The analysis of \nus\ observations of \obj\ from August 2021 revealed that the source was in a low-flux X-ray state, evident in energies up to $\sim$40 keV. This period, marked by the second \nus\ campaign targeting \obj, is clearly characterised by a low-X-ray state, as seen in the long-term 0.3-10\,keV light curve of the blazar.

These \nus\ observations also revealed a concave spectral shape with a notable upturn of around 25 keV. Such an upturn in the X-ray range had not been reported before, neither in previous \nus\ observations nor in \textit{Swift}-BAT data. Furthermore, spectral modelling by \cite{Kaufmann} suggests that the X-ray emission of \obj\ extends up to 100\,keV without any cut-off or spectral upturn.

Here, we report a distinct concave spectral shape observed in all three observations from August 2021, with the upturn located at 25–26 keV. Despite temporal variability in these observations, with changes in the 3–10 keV flux of approximately 20$\%$, the position of the spectral upturn remains consistent within the parameter uncertainties.

To explore the origin of the X-ray excess above 25\,keV, several scenarios can be considered. One possible explanation for the concave X-ray spectrum is the presence of a spectral crossing point in this energy range, where the synchrotron and inverse Compton components converge. Interestingly, similar spectral upturns, linked to the presence of synchrotron and inverse Compton components, have been detected in the 0.3-10\,keV X-ray band for several IBL-type blazars \citep[see e.g.][]{Wierzcholska_swift}. The authors also suggested that for some IBL-type blazars, the spectral upturn may be located in the X-ray domain, but above 10\,keV.

For HBL-type sources, a concave X-ray spectrum has been observed in only a few cases, such as PKS\,2155-304 and Mrk\,421 (\citealt{Zhang08} and \citealt{Kataoka}, respectively). In the latter case, the spectral upturn was detected during the \nus\ observation of the source's low state, but was absent in high-state observations of Mrk\,421.

Alternatively, the three-bump SED of blazars, as indicated by the \nus\ spectrum of \obj, may be explained by a photohadronic origin. For HBL-type blazars, the energy range of 40 keV to 40 MeV is where a third SED bump might be visible. It is well established that both electrons and protons can be accelerated to relativistic energies and may contribute to the broadband emission observed \citep[see e.g.][]{Biermann, Sironi}. In the leptohadronic scenario, the low-energy component of the SED is explained by synchrotron emission from relativistic electrons, whereas the high-energy component involves interactions of relativistic protons in the jets. \cite{Petropoulou} demonstrated that Bethe-Heitler pair production and photopion production play significant roles in this process. The authors showed that when relativistic protons interact with synchrotron photons in blazars, leading to gamma-ray production through photopion processes, the SED is affected by two key features: PeV neutrino emission and a third SED bump located in the keV–MeV energy range. A natural implication of this scenario and the leptohadronic origin of blazar emission is the potential for high-energy neutrino emission from blazars exhibiting a three-bump SED.

Having observational constraints, the best way to distinguish between different possible scenarios describing physical processes responsible for a broadband emission is a modelling. 
However, in the case of \obj, due to the weakness of high-energy $\gamma$-ray emission, it is not possible to constrain the Fermi-LAT spectrum using a few days or even a few weeks of data. 
Thus, a discussion of this energy regime in terms of the low state of \obj\ is not possible \citep[see e.g.][]{Cologna}.

Considering the discussion provided by \cite{Wierzcholska_host} on the host galaxy, spectral curvature, absorption, and ultraviolet excess features in the broadband SED of \obj,  we argue that the feature detected in the low-state spectrum in this work is more likely attributed to photohadronic processes.
That is also supported by the consideration of signatures of the Bethe-Heitler emission in the blazars' broadband SEDs provided by \cite{Petropoulou}.
Assuming this scenario, \obj\ should also be considered a potential source of high-energy neutrino emission.

Furthermore, it should be noted that in previous X-ray observations of \obj\ conducted by \nus\ and \textit{Swift}-BAT, the upturn was not observed. This can be explained by the higher X-ray state of the source compared to the August 2021 observations discussed in this work.

\begin{acknowledgements}
The project is co-financed by the Polish National Agency for Academic Exchange.
The authors gratefully acknowledge the Polish high-performance computing infrastructure PLGrid (HPC Centre: ACK Cyfronet AGH) for providing computer facilities and support within computational grant no. \text{PLG/2024/017925}. 
This research has made use of the NuSTAR Data Analysis Software (NuSTARDAS) jointly developed by the ASI Space Science Data Center (SSDC, Italy) and the California Institute of Technology (Caltech, USA).
\end{acknowledgements}

\bibliographystyle{aa} 
\bibliography{references.bib}

\end{document}